\documentclass[12pt]{article}
\begin{document}
\baselineskip 18pt
\newcommand{\Tr}{{\mathrm{Tr}\,}}
\newcommand{\Dirac}{/\!\!\!\!D}
\newcommand{\beq}{\begin{equation}}
\newcommand{\eeq}[1]{\label{#1}\end{equation}}
\newcommand{\bea}{\begin{eqnarray}}
\newcommand{\eea}[1]{\label{#1}\end{eqnarray}}
\renewcommand{\Re}{{\mathrm{Re}\,}}
\renewcommand{\Im}{{\mathrm{Im}\,}}
\begin{titlepage}
\hfill  CERN-TH/99-89 NYU-TH/99/03/01 hep-th/9903241
\begin{center}
\hfill
\vskip .4in
{\large\bf $AdS$ Superalgebras with Brane Charges}
\end{center}
\vskip .4in
\begin{center}
{\large S. Ferrara$^a$, M. Porrati$^{a,b}$\footnotemark}
\footnotetext{e-mail: sergio.ferrara@cern.ch,
massimo.porrati@nyu.edu}
\vskip .1in
(a){\em Theory Division CERN, Ch 1211 Geneva 23, Switzerland}
\vskip .1in
(b){\em Department of Physics, NYU, 4 Washington Pl.,
New York, NY 10003, USA\footnotemark}
\footnotetext{Permanent Address}
\end{center}
\vskip .4in
\begin{center} {\bf ABSTRACT} \end{center}
\begin{quotation}
\noindent
We consider the inclusion of brane charges in $AdS_5$
superalgebras that contain
the maximal central extension of the super-Poincar\'e algebra on $\partial
AdS_5$.
For theories with $N$ supersymmetries on the boundary, the maximal extension
is $OSp(1/8N,R)$, which contains the group
$Sp(8N,R)\supset U(2N,2N) \supset SU(2,2)\times U(N)$ as extension of the
conformal group. An ``intermediate'' extension to $U(2N,2N/1)$ is also
discussed, as well as the inclusion of brane charges in $AdS_7$ and
$AdS_4$ superalgebras.
BPS conditions in the presence of brane charges are studied in some details.
\end{quotation}
\vfill
CERN-TH/99-89 \\
March 1999
\end{titlepage}
\eject
\noindent
\section{Introduction}
It is well known that the classification of superalgebras, containing the
anti-de Sitter --i.e. conformal-- superalgebra has a barrier at dimension
$D=7$~\cite{n}. This result is based on the assumption that the bosonic
subalgebra of the superalgebra is the ``direct product'' of the conformal
algebra $O(D-1,2)$ with an internal symmetry group $G$.
This classification can be viewed, essentially, as an extension of the
Haag-Lopuszanski-Sohnius theorem~\cite{hls} for $D>4$.

On the other hand, in dynamical theories with extended objects, it is already
known that the super-Poincar\'e algebras do include, in any dimension,
``central'' charges~\cite{agit,t1,b,t2,fp}
which at first sight violate the above assumption.

It is then natural to consider the same generalization when such charges of
rank $p$ are introduced in the AdS superalgebra rather than in its Poincar\'e
counterpart.
Such extension has been studied in the literature with the goal of
constructing an eleven-dimensional theory in AdS space~\cite{h,g},
or a conformal theory in ten dimensions~\cite{t2}.

The result of these investigations is that the conformal extension of the
$D=10$ $N=1$ Poincar\'e superalgebra is $OSp(1/32,R)$, in which
$Sp(32,R)\supset O(10,2)$.

In the present paper, we consider a similar extension in the context of
$AdS_5$ supergravity. The difference here is that $AdS_5$ superalgebras
already exist. In the absence of $p$-brane charge, they correspond to the
usual superconformal algebras $U(2,2/N)$, which occur in the classification of
Haag, Lopuszanski and Sohnius (HLS)~\cite{hls}.
We consider here the $AdS_5$ superalgebra in the presence of AdS $p$-branes,
and we show that, for any $N$-extended supergravity, such algebra is
$OSp(1/8N)$, with the conformal group $O(4,2)\sim SU(2,2)$ embedded as follows
in the real symplectic group $Sp(8N,R)$~\cite{gil}:
\beq
Sp(8N,R)\supset U(2N,2N) \supset SU(2,2)\times U(N).
\eeq{1}
Extensions of conformal superalgebras in $D=5$ have been considered also
in ref.~\cite{cgmp}, where worldsheet superalgebras for $D5$-branes in
an $AdS_5$ background were proposed. Ref.~\cite{cgmp} shows, among other
things, that, in the presence of $p$-brane charges, the world-sheet
superconformal group of a $D3$-brane in $AdS_5$, $U(2,2/4)$, is
extended to $OSp(1/32)$.

This paper is organized as follows: in Section 2, we consider the standard
superalgebras in the $5$-$D$ Minkowski space, $M_5$, and in $AdS_5$,
and view them as the
starting blocks for further investigations.
In Section 3 we consider the $OSp(1/8N)$ algebras as algebras in $AdS_5$ in the
presence of AdS $p$-branes. In Section 4 we give a general, algebraic
analysis of the BPS condition in $AdS_5$, and, in particular, we
study the pattern of $R$-parity breaking induced by BPS $p$-branes.
In Section 5 we present an extension of the analysis performed
in Sections 3 and 4 to $AdS_7$.
Section 6 contains a brief description of an additional ``intermediate''
conformal extension of the 5-$D$ super-Poincar\'e algebra, a comment on
the uniqueness of such extensions, and a brief analysis of $Ads_4$
superalgebras and their 1/2 BPS brane configurations.
\section{Maximal Central Extensions of Poincar\'e Superalgebras}
The maximal central extension of the Poincar\'e superalgebra with
$n$ spinorial components of the supersymmetry charges gives an algebra
with $n(n+1)/2$ bosonic central charges~\footnotemark,
including the space-time translations.
\footnotetext{More precisely, they are bosonic central charges of the
supertranslation algebra, which is a subalgebra of the super-Poincar\'e
algebra.}
Examples of such extensions are the $N=1$ superalgebra in $D=11$
dimensions,
in the presence of two- and five-brane charges, and
the IIA and IIB algebras in ten dimensions, in the presence of NS and R brane
charges~\cite{agit,t1,b,t2}.

When the space-time dimension is sufficiently low, the maximal central
extensions include also BPS domain walls and BPS instantons, as it becomes
obvious if one regards such algebras obtained by dimensional reduction.
It is relevant to this paper to recall the central extensions in dimensions
$D=4,5$, because they will play an important role when the analogous
five-dimensional superalgebra will be considered in $AdS_5$, and $M_4$
will be interpreted as its boundary.

The $N$-extended Poincar\'e superalgebra in $D=5$, with maximal central
extension, has a $USp(2N)$ R-symmetry~\cite{crem}, and it reads
\beq
\{ Q^A_\alpha, Q^B_\beta\} = (\gamma^\mu C)_{\alpha\beta} P_\mu \Omega^{AB}
+ (\gamma^\mu C)_{\alpha\beta}Z^{o\,[AB]}_\mu +
C_{\alpha\beta}Z^{[AB]} + (\gamma^{\mu\nu} C)_{\alpha\beta}Z_{\mu\nu}^{(AB)},
\eeq{2}
where $Z^{o\,[AB]}_\mu$, $Z^{[AB]}$ are in the antisymmetric of $USp(2N)$
($Z^{o\,[AB]}_\mu$ is also symplectic-traceless:
$\Omega_{AB} Z^{o\,[AB]}_\mu =0$) and
$Z_{\mu\nu}^{(AB)}$ is in the adjoint of
$USp(2N)$.

The standard HLS~\cite{hls} algebra is obtained by setting $Z^{o\,[AB]}_\mu=
Z_{\mu\nu}^{(AB)}=0$. These charges come from strings and membranes.
\section{Anti-de Sitter and Conformal Superalgebras}
The Anti-de Sitter superalgebra is a modification of the Poincar\'e
superalgebra given in Eq.~(\ref{2}),
where $P_\mu$, $M_{\mu\nu}$ span the algebra of
$O(4,2)$, and generators of $U(N)$ are included.

This is the $AdS_5$ superalgebra $U(2,2/N)$. This superalgebra can be
formally obtained by decomposing $USp(2N)\rightarrow SU(N)\times U(1)$ in the
former algebra:
\bea
\{ Q^A_\alpha, Q_{B\,\beta}\}&=& (\gamma^\mu C)_{\alpha\beta} P_\mu \delta^A_B
+ (\gamma^{\mu\nu} C)_{\alpha\beta}M_{\mu\nu}\delta^A_B +
(\gamma^{\mu\nu} C)_{\alpha\beta} Z^{o\,A}_{\mu\nu\, B} +
(\gamma^{\mu} C)_{\alpha\beta}Z^{o\,A}_{\mu\, B} +C_{\alpha\beta} U^A_B,
\nonumber \\ && \label{3}\\
\{ Q^A_\alpha, Q^B_\beta\}&=&  (\gamma^\mu C)_{\alpha\beta}Z^{[AB]}_\mu
+ C_{\alpha\beta} Z^{[AB]} + (\gamma^{\mu\nu} C)_{\alpha\beta}
Z_{\mu\nu}^{(AB)},\;\;\; c.c., \;\;\; \mu,\nu=0,..,4 .
\eea{4}
Setting to zero all bosonic generators except  $P_\mu$, $M_{\mu\nu}$ and
$U^A_B$, and promoting them to the (non-commutative) generators of
the $SU(2,2)\times U(N)$ Lie algebra, with the fermionic generators in the
$(4,N)+(\bar{4},\bar{N})$ representation, the algebra becomes
\bea
\{ Q^A_\alpha, Q_{B\,\beta}\}&=& (\gamma^\mu C)_{\alpha\beta} P_\mu \delta^A_B
+ (\gamma^{\mu\nu} C)_{\alpha\beta}M_{\mu\nu} + C_{\alpha\beta}U^A_B,
\label{5}\\
\{ Q^A_\alpha, Q^B_\beta\}&=&\{ Q_{A\,\alpha}, Q_{B\,\beta}\}=0.
\eea{6}
This is the standard $AdS_5$ superalgebra considered in the literature.
If realized on the four-dimensional boundary, it corresponds to the HLS
superconformal algebra in $D=4$, which is the conformal extension of the
Poincar\'e superalgebra without central charges.

Let us now consider whether an $AdS_5$ superalgebra exists with non-vanishing
$Z$ generators. The $Z$ should correspond somehow
to brane charges in $AdS_5$.

The $AdS_5$ extension of the Poincar\'e superalgebra with charges given in
Eqs.~(\ref{3},\ref{4}) is immediate. Namely, the $Z$ generators in
Eqs.~(\ref{3},\ref{4}) complete the superalgebra $OSp(1/8N,R)$. The
$SU(2,2)\times U(N)$ generators are embedded as follows in $Sp(8N,R)$:
\beq
Sp(8N,R)\rightarrow U(2N,2N) \rightarrow SU(2,2)\times U(N).
\eeq{7}
As it is well known~\cite{g}, the $Sp(8N,R)$ algebra has a three-grading with
respect to the ``dilation'' generator, $R$ in the decomposition
$SU(2,2)\rightarrow SL(2,C)\times R$. Here, $SL(2,C)$ is the Lorentz group of
the boundary of $AdS_5$. Indeed,
\beq
{\cal L}_{Sp(8N)}= {\cal L}^1 + {\cal L}^0 + {\cal L}^{-1},
\eeq{8}
where ${\cal L}^1$ contains $P_\mu$ and all (dimension-1) central charges
of the $4$-$D$ super-Poincar\'e algebra. ${\cal L}^{-1}$ contains $K_\mu$
and all (dimension $-1$) special-conformal  central charges, while
\beq
{\cal L}^0= SL(4N,R)\times R
\eeq{9}
is the Lie algebra which contains, among others, the generators of
the Lorentz group on $\partial AdS_5$, and $U(N)$~\cite{gil}:
\beq
SL(4N,R)\rightarrow SL(2N,C)\times U(1)\rightarrow SL(2,C) \times SU(N)\times
U(1).
\eeq{9a}
The $OSp(1/8N)$ superalgebra has a 5-grading~\cite{g,g2,g3},
in which the ${\cal L}^{\pm 1}$
subalgebras, in the symmetric representation of $SL(4N)$,
are completed with the $8N$-dimensional
fundamental representation of
$Sp(8N,R)$, which splits under $SL(2,C)\times SU(N)\times U(1)$ as:
\beq
8N \rightarrow (1/2,0,N)^{1/2} + (0,1/2,N)^{1/2} + (0,1/2,\bar{N})^{-1/2}
+ (1/2,0,\bar{N})^{-1/2}.
\eeq{10}
This splitting corresponds to writing the $AdS_5$ spinor $Q_\alpha^A$ as
$(Q^A_\alpha , \bar{S}^A_{\dot{\alpha}})$, and $Q_{\alpha\,A}$ as
$(\bar{Q}_{\dot{\alpha}\, A} , S_{\alpha\, A})$. The spinor charges
$(Q_\alpha^A,\bar{Q}_{\dot{\alpha}\, A})$,
together with ${\cal L}^1$, form the maximal
central extension of the super-Poincar\'e algebra in $D=4$, as found in
ref.~\cite{fp}. The $(S_{\beta\, A},\bar{S}^A_{\dot{\beta}})$, together with
${\cal L}^{-1}$, form an isomorphic algebra, with the substitution
$P_\mu\rightarrow K_\mu $ and $Z\rightarrow Z_S$. The generators in
${\cal L}^0$ appear in the mixed anti-commutators $\{Q,S\}$, as it follows
from the general structure of the grading:
\bea
{\cal SP}^{+}&:& \begin{array}{lll} \{ Q^A_\alpha, \bar{Q}_{\dot{\beta}\,B}\}
&=&
\sigma_{\alpha\dot{\beta}}^\mu P_\mu \delta^A_B +
Z^A_{\alpha\dot{\beta}\,B},
\;\;\; (Z^A_{\alpha\dot{\beta}\, A}=0),\;\;\; \mu=0,..3,
\\ \{Q^A_\alpha, Q^B_\beta\}&=&
\epsilon_{\alpha\beta} Z^{[AB]} + Z^{(AB)}_{\alpha\beta},\\ && c.c.,
\end{array} \nonumber\\
{\cal SC}^{-}&:& \begin{array}{lll} \{ S_{\alpha\,A}, \bar{S}^B_{\dot{\beta}}\}
&=&
\sigma_{\alpha\dot{\beta}}^\mu K_\mu \delta^B_A +
Z^B_{S\,\alpha\dot{\beta}\,A},
\;\;\; (Z^A_{S\,\alpha\dot{\beta}\, A}=0),\\ \{S_{\alpha\,A}, S_{\beta\,B}\}
&=&
\epsilon_{\alpha\beta} Z_{S\,[AB]} + Z_{S\,\alpha\beta\,(AB)},\\ && c.c.,
\end{array} \nonumber\\
{\cal L}^0 &:& \begin{array}{lll} \{Q^A_\alpha, S_{\beta\,B}\} &=&
[\epsilon_{\alpha\beta}(D+iU) +M_{\alpha\beta}]\delta^A_B +
\epsilon_{\alpha\beta}U^{o\,A}_B + U^{o\,A}_{\alpha\beta\,B}\;\;\;
(\mathrm{traceless}), \\
\{Q^A_{\alpha}, \bar{S}^B_{\dot{\beta}}\}&=& W_{\alpha\dot{\beta}}^{[AB]}
+ W_{\alpha\dot{\beta}}^{(AB)},\\ && c.c.,\end{array} \nonumber\\
{\cal L}_{OSp(1/8N)}&=& {\cal SP}^+ + {\cal L}^0   + {\cal SC}^- .
\eea{11}
The total number of generators of $Sp(8N,R)$ is $4N(8N+1)$, which splits, in
this decomposition, as
\bea
&&(\mathrm{Sym} \, GL(4N))^+ + (\mathrm{Adj}\, GL(4N))
+ (\mathrm{Sym} \, GL(4N))^-\\ && \mathrm{dim}\, {\cal P}^+= \mathrm{dim}\,
{\cal C}^-= 8N^2 +2N,\\ && \mathrm{dim}\, {\cal L}^0 = 16N^2.
\eea{12}
Note that in the
usual conformal extension of the non-centrally extended ${\cal SP}$,
that we call ${\cal P}_0$,
\beq
{\cal L}_{U(2,2/N)}={\cal SP}^+_0 + {\cal L}^0_0 + {\cal SC}^-_0,
\eeq{13}
where $\mathrm{dim}\, {\cal P}_0^+=\mathrm{dim}\, {\cal C}^-_0=4$,
$ \mathrm{dim}\, {\cal L}_0^0=15+N^2$.
\section{BPS States in $AdS_5$ and R-Symmetry Breaking}
States that preserve some of the supersymmetries of
the $D=4$ $N$-extended
super-Poincar\'e algebra ${\cal SP}^+$ can be point-like or extended.
These states preserve only subgroups of the R-symmetry $U(N)$, and their
breaking pattern can be analyzed in pure algebraic terms. This analysis
agrees with previous studies of branes in $AdS_5$~\cite{cgmp,w2} in all known
cases, but also predicts general patterns of R-symmetry breaking.
It should be emphasized that our analysis is purely in terms of Poincar\'e
multiplets, and that we do not know whether all breaking patterns we find here
are effectively realized in the AdS/CFT correspondence~\cite{malda,gkp,w1,w2}.

The identification of brane charges with the central charges of the
super-Poincar\'e algebra in Eq.~(\ref{2}) is well established in flat space.
The corresponding identification of brane charges with some bosonic
genrators of $OSp(1/8N)$ should hold in $AdS_5$~\cite{cgmp,w2}. This
correspondence
was established explicitly in~\cite{cgmp} for a brane charge appearing
in the world-volume superalgebra of a $D5$-brane in $AdS_5$.

Let us consider, in particular, the case $N=4$, and let us start our analysis
with point-like states (monopoles and dyons). They are associated with
a scalar central charge, that we called $Z^{[AB]}$. It can always be
put in the form
\beq
Z^{[AB]}=\left(\begin{array}{ll}
\lambda_1 \epsilon & \\ &\lambda_2 \epsilon \end{array} \right),
\eeq{14}
with $\epsilon$ the $2\times 2$ antisymmetric matrix.
When $\lambda_1=\lambda_2$ we have a $1/2$ BPS state, and the R-symmetry
$SU(4)$ is broken to $USp(4)\sim O(5)$.
When $\lambda_1\neq \lambda_2$ one has
$1/4$ BPS states, and $SU(4)\rightarrow USp(2)\times USp(2)\sim O(3)\times
O(3)$.

String BPS states are charged under the vector charge $Z^A_{\mu\,B}$.
Let us consider a string oriented along one of the coordinate axis. The only
nonzero component of $Z^A_{\mu\,B}$ is, say  $Z^A_{1\,B}$ that can be
brought to the standard form:
\beq
Z^A_{\mu\,B}=\left(\begin{array}{llll}
\lambda_1 & & &\\ &\lambda_2 &&\\ && \lambda_3 & \\ &&& -\lambda_1 -\lambda_2
-\lambda_3 \end{array} \right).
\eeq{15}
By looking at the $SU(4)\sim O(6)$ subgroup left invariant by this matrix
one can easily find all possible R-symmetry breaking patterns.
Specifically,
$1/2$ BPS states have $\lambda_1=\lambda_2=-\lambda_3$; this means that the
R-symmetry is broken to $SU(2)\times SU(2) \times U(1)$, or
\beq
O(6)\rightarrow O(4)\times O(2).
\eeq{16}
$1/4$ BPS states have $\lambda_1=\lambda_2$, $\lambda_3\neq \lambda_1$.
The R-symmetry breaking is in this case
\beq
O(6)\rightarrow SU(2)\times U(1)^2.
\eeq{17}
When $\lambda_1=\lambda_2=\lambda_3$ the string preserves $3/8$ of the
original supersymmetries, while
\beq
O(6)\rightarrow SU(3)\times U(1).
\eeq{18}
Finally, for generic $\lambda_i$, one finds a $1/8$ BPS state preserving
only a $U(1)^3$ subgroup of the R-symmetry.

Finally, 2-branes are sources for the antisymmetric-tensor charge
$Z_{\mu\nu}^{(AB)}$, belonging to the symmetric representation of $SU(4)$.
By choosing a 2-brane configuration oriented along two coordinate axis,
the only nonzero charge is, say, $Z_{12}^{(AB)}$. It can be brought into
the standard form~\cite{zu}
\beq
Z_{12}^{(AB)}=\left(\begin{array}{llll}
\lambda_1 && & \\&\lambda_2 && \\ && \lambda_3 & \\ &&& \lambda_4
\end{array} \right), \;\;\; \lambda_i \geq 0.
\eeq{19}
In this case, $1/2$ BPS states correspond to all $\lambda_i=\lambda$, and
the R-symmetry breaking pattern is
\beq
SU(4)\rightarrow O(4)=O(3)\times O(3).
\eeq{1/2}
$1/4$ BPS states arise in two cases. First, when $\lambda_1=\lambda_2$,
$\lambda_3=\lambda_4\neq \lambda_1$. The residual
R-symmetry is, in this case
\beq
SU(4)\rightarrow O(2)\times O(2).
\eeq{1/4a}
In the second case, $\lambda_1=\lambda_2$, $\lambda_3\neq\lambda_4$ and
neither $\lambda_3$ nor $\lambda_4$ equal $\lambda_1$. The R-symmetry
breaks as
\beq
SU(4) \rightarrow O(2).
\eeq{1/4b}
If $\lambda_1=\lambda_2=\lambda_3\neq \lambda_4$, we have $3/8$ BPS states
with R-symmetry
\beq
SU(4) \rightarrow O(3).
\eeq{3/8}
Finally, if all $\lambda_i$ are different, we have $1/8$ BPS states,
which completely break R-symmetry.

The 1/2
BPS states on the boundary can be thought of as ``singletons'' since they
have multiplicity $2^N$ ($N=4$ in our case), if regarded as $N$-extended
$4$-$D$ Poincar\'e multiplets.

We call $s$-singleton the usual singleton associated to a massless particle
on the boundary, while we call $p$-singleton a state associated to a
$p$-brane on the boundary. Thus, ``photons'' are $s$-singletons, monopoles or
dyons are 0-singletons etc.
BPS states propagating in the bulk are ``bound states'' of $p$-singletons,
since they have multiplicity $2^{2N}$.

To summarize, a $p$-singleton breaks the original $O(6)$ R-symmetry to
$O(5-p)\times O(p+1)$, while the $s$-singleton corresponds to $p=-1$ in
this formula.
\section{Extension to $AdS_7$ Superalgebras}
The analysis performed in the previous two Sections can be extended to the
$AdS_7$ case, based on the conformal algebra
$O^*(8) \sim O(6,2)$.
This case is related to the conjectured duality~\cite{kvp,malda}
between M-theory on $AdS_7$ and superconformal field theories in 6-$D$.

Here, maximal or reduced supersymmetry correspond to the
$OSp(8^*/2N)$ superalgebras, with $N=2,1$, respectively.

The contraction of these algebras gives the non-maximally centrally extended
supertranslation algebra in $D=7$.

The maximal central extension of the supertranslation algebra, on the other
hand, reads
\bea
\{Q_\alpha^A,Q_\beta^B\} &=& (\gamma^\mu C)_{\alpha\beta}P_\mu \Omega^{AB} +
(\gamma^{\mu\nu}C)_{\alpha\beta} M_{\mu\nu}\Omega^{AB} +
(\gamma^\mu C)_{\alpha\beta}Z_\mu^{o\,[AB]} +
\nonumber \\
&& + (\gamma^{\mu\nu} C)_{\alpha\beta}Z_{\mu\nu}^{o\,[AB]} +
C_{\alpha\beta}U^{(AB)} +
(\gamma^{\mu\nu\rho} C)_{\alpha\beta}Z_{\mu\nu\rho}^{(AB)}, \nonumber \\
&& A,B=1,..,4,\;\;\; \alpha,\beta=1,..,8,\;\;\; \mu,..=0,..,6.
\eea{c7}
Its extension, containing the $AdS_7$ isometry algebra $O(6,2)$, and the
R-symmetry algebra $USp(2N)$, is the superalgebra
$OSp(1/16N)$, with $N=2,1$ respectively. The $O^*(8)$ algebra appears in the
decomposition~\footnotemark
\beq
Sp(16N,R) \rightarrow USp(2N)\times O^*(8).
\eeq{o*}
\footnotetext{This embedding is obtained by writing the symplectic metric
as $\Omega_{16N}=\Omega_{2N}\otimes I_{8}$. Then, the
generators of $Sp(16N,R)$ of factorized form $A=a\otimes I_{8}$ or
$A=I_{2N} \otimes b$, with $a$, $b$ $2N\times 2N$ and $8\times 8$ matrices,
respectively, form manifestly a subgroup. Here
$I_{n}$ is the $n\times n$ identity matrix and $\Omega_{n}$ is the canonical
$n\times n$ symplectic metric. The symplectic condition on the $Sp(16N,R)$
generators is $\Omega A^T \Omega =A$. The reality condition can be written in
different forms. By choosing $A=A^*$, the subgroup of factorized matrices is
$Sp(2N,R)\times SO(8)$. By choosing instead $A=\Sigma A^* \Sigma$, with
$\Sigma=\Omega_{2N}\otimes \Omega_{8}$, the subgroup is
$USp(2N)\times O^*(8)$ --since $a$ obeys the constraints $a^\dagger =-a$,
$\Omega_{2N}a^T\Omega_{2N}=a$, that define the $USp(2N)$ algebra, while $b$ is
antisymmetric, $b^T=-b$, and quaternionic, $\Omega_8 b^* \Omega_8=-b$. The
latter constraints define the generators of $O(4,Q)\sim O^*(8)$}

The $OSp(1/16N)$ superalgebra can be written in a manifestly $O(6,2)$-invariant
form as follows:
\bea
\{Q_\alpha^A,Q_\beta^B\}^+ &=& (\gamma^{\mu\nu} C)^+_{\alpha\beta}
M_{\mu\nu} \Omega^{AB} +
(\gamma^{\mu\nu} C)^+_{\alpha\beta}Z^{o\,[AB]}_{\mu\nu} +
(\gamma^{\mu\nu\rho\sigma} C)^+_{\alpha\beta}
Z^{+\,(AB)}_{\mu\nu\rho\sigma} + C^+_{\alpha\beta} U^{(AB)}, \nonumber\\
&& \alpha,\beta=1,..,8,\;\;\; \mu..=0,..,7.
\eea{mancov}
The subindex $+$ denotes the 8-$D$ chiral projection, and
$Z^{+\,(AB)}_{\mu\nu\rho\sigma}$ is self-dual.

This superalgebra reduces to $OSp(8^*/2N)$ by setting to zero the generators
$Z^{o\,[AB]}_{\mu\nu}$ and $Z^{+\,(AB)}_{\mu\nu\rho\sigma}$.

Notice that for $N=2$, the embedding~(\ref{o*})
of $O^*(8)$ in $Sp(32,R)$ is not unique. Indeed,it differs from the embedding
obtained by dimensional reduction of $OSp(1/32)$, interpreted
as $AdS_{11}$ superalgebra~\cite{t2}.
In this case, the superalgebra reads
\bea
\{Q_\alpha,Q_\beta\}&=& (\gamma^{MN}C)_{\alpha\beta}L_{MN}
+ (\gamma^{MNPQRS}C)_{\alpha\beta}
T^+_{MNPQRS}, \nonumber \\
&& \alpha..=1,..,32,\;\;\; M..=0,..,11,
\eea{town}
where $Q_\alpha$ is a 12-$D$ Majorana-Weyl spinor
and $T^+_{MNPQRS}$ is self-dual.
By splitting the index $M=\mu,i$ ($\mu=0,..,7$, $i=1,..,4$) it is easy to
see that the above superalgebra contains two subalgebras $OSp(1/16)_{L,R}$,
where $Q_\alpha$ splits as $32 =16_L +16_R$. The bosonic part
of the superalgebra breaks as
\beq
Sp(32,R) \rightarrow
SO(10,2) \rightarrow SO(4)\times SO(6,2),
\eeq{11demb}
where $SO(4)=SO(3)_L\times SO(3)_R$, and $SO(6,2)=O^*(8)_L+O^*(8)_R$.
The 528 generators of $Sp(32,R)$ are in this case arranged as two copies of
$Sp(16,R)$, and 256 additional generators, $Z_{\mu i}$, $Z_{\mu\nu\rho ij}$
($\mu..=0,..,7$, $i,j=1,..,4$). These new generators are elements of the
coset $Sp(32,R)/Sp(16,R)_L\times Sp(16,R)_R$.
They are contained in the mixed anticommutators
$\{16_L,16_R\}$.

The spinor charges decompose as follows as representations of
$SO(6,2)\times SO(4)$:
\beq
32 = (8_L,2) + (8_R, 2');
\eeq{88}
as representations of $O^*(8)_L\times O(3)_L\times O^*(8)_R \times O(3)_R
\subset Sp(16,R)_L\times Sp(16,R)_R \subset Sp(32,R)$, the spinor charges
decompose instead as:
\beq
32 = (8_L,2_L,1,1) + (1,1,8_R, 2_R).
\eeq{8383}
If one drops the above 256 generators one finds two copies of
the $OSp(1/16)$ superalgebra. By further truncation one gets
$OSp(8^*/2)_L\times OSp(8^*/2)_R$. Therefore, the two embeddings only coincide
when the generators are restricted to a $(1,0)$ superconformal extension.

The superalgebra in Eq.~(\ref{mancov}),
interpreted on $\partial AdS_7$, gives a conformal extension
of the maximally centrally extended supertranslation algebra of type
$(2,0)$ and $(1,0)$, respectively.

For $N=2$, this is the $OSp(1/32,R)$ algebra given in Eqs~(\ref{c7},
\ref{mancov}),
now written in 6-$D$ notations:
\bea
\{Q_\alpha^A,Q_\beta^B\}^+ &=& (\gamma^\mu C)^+_{\alpha\beta}
P_\mu \Omega^{AB} +
(\gamma^\mu C)^+_{\alpha\beta}Z_\mu^{o\,[AB]} +
(\gamma^{\mu\nu\rho} C)^+_{\alpha\beta}Z_{\mu\nu\rho}^{(AB)}
, \nonumber \\
\{S_\alpha^A,S_\beta^B\}^- &=& (\gamma^\mu C)^-_{\alpha\beta}
K_\mu \Omega^{AB} +
(\gamma^\mu C)^-_{\alpha\beta}Z_{S\,\mu}^{o\,[AB]} +
(\gamma^{\mu\nu\rho} C)^-_{\alpha\beta}
Z_{S\,\mu\nu\rho}^{+\,(AB)}, \nonumber \\
\{Q_\alpha^A,S_\beta^B\} &=& C_{\alpha\beta} U^{o\,AB} +
(\gamma^{\mu\nu} C)_{\alpha\beta}M_{\mu\nu}\Omega^{AB} +
C_{\alpha\beta} D \Omega^{AB}
+ (\gamma^{\mu\nu} C)_{\alpha\beta}Z_{\mu\nu}^{o\,AB}, \nonumber \\
&& \alpha,\beta=1,..,4,\;\;\; \mu..=0,..,5.
\eea{32}
The subindex $+$ is the chiral projector in 6-$D$ and
$Z_{S\,\mu\nu\rho}^{+\,(AB)}$ is self-dual.
The BPS analysis with this boundary algebra can be done using the methods
explained in Section 4. In particular, looking at the super-Poincar\'e
algebra, we find the following patterns of R-symmetry breaking:
Strings are 1/2 BPS carrying
a nonzero $Z^{o\,[AB]}$ charge. With an appropriate choice
of basis it can be cast in the form
\beq
Z^{o\,[AB]}= \left( \begin{array}{ll} \lambda \epsilon &  \\ & -\lambda
\epsilon \end{array} \right).
\eeq{7st}
This matrix breaks $USp(4)\sim O(5) \rightarrow USp(2)\times USp(2) \sim O(4)$.

Three-branes carry a nonzero $Z^{(AB)}$ charge. It can be reduced to the
following form:
\beq
Z^{(AB)}=\left( \begin{array}{ll} \lambda_1\delta  &  \\ & \lambda_2\delta
\end{array} \right),
\eeq{3b7}
where $\delta$ is the $2\times 2$ identity matrix.
When $\lambda_1=\lambda_2$, the corresponding state is 1/2 BPS, and the
R-symmetry breaks as follows: $USp(4)\sim O(5) \rightarrow O(3) \times O(2)$.
When $\lambda_1\neq \lambda_2$, the state is $1/4$ BPS, and the $USp(4)$
R-symmetry breaks to $O(2)\times O(2)$. Particles and membranes have the same
R-symmetry breaking pattern as three-branes and strings.
These results agree with the explicit analysis performed in
refs.~\cite{lm,aky}
\section{Other Algebras and Uniqueness of the Extension}
Let us conclude with a few additional remarks.

First, let us point out that
here is a superalgebra that is intermediate between that in
Eqs.~(\ref{3},\ref{4}) and that in Eqs.~(\ref{5},\ref{6}). This is the
$U(2N,2N/1)$ superalgebra. It is obtained by setting to zero the right-hand
side of Eq.~(\ref{4}), but keeping all terms in the r.h.s. of Eq.~(\ref{3}).

All these algebras can be written in a manifestly $O(4,2)$-invariant
notation using the following decompostition, where $A$ and $S$ denote
symmetrization and antisymmetrization, respectively ($\mu,\nu=0,..5$):
\bea
[(4,N)\times (4,N)]_S &=& [6, (N\times N)_A] + [10,(N\times N)_S], \nonumber \\
(4,N)\times (\bar{4},\bar{N}) &=& (1,1) +(15,1) + (15, N^2-1)+ (1,N^2-1),
\nonumber \\
\{ Q^A,Q^B\} &=& (\gamma^\mu C)^+T^{[AB]}_\mu + (\gamma^{\mu\nu\rho}C)^+
T^{+\,(AB)}_{\mu\nu\rho}, \;\;\; c.c. \nonumber \\
\{ Q^A,Q_B\} &=& (\gamma^{\mu\nu} C)M_{\mu\nu} \delta^A_B + CT^A_B +
(\gamma^{\mu\nu} C)T^{o\,A}_{\mu\nu \, B}.
\eea{20}
Here $+$ is the $6$-$D$ chiral projection and $T^{+\,(AB)}_{\mu\nu\rho}$ is
self-dual.
The space-time conformal spinors are identified here with the $(4,\bar{4})$
of $SU(2,2)$. The $U(2N,2N/1)$ superalgebra is obtained by setting
$\{Q^A,Q^B\}=0$, while the $U(2,2/N)$ superalgebra is obtained by setting
$\{Q^A,Q^B\}=0$ and $ T^{o\,A}_{\mu\nu \, B}=0$.

We must also point out that $AdS_5$ algebras with brane charges
clearly violate the Coleman-Mandula theorem~\cite{cm}. This should imply
that they cannot be realized as symmetries of a local world-sheet
theory. In spite of this, the brane charges studied in ref.~\cite{cgmp}
are topological charges appearing in the (local) Born-Infeld action of
the $D$-brane. The meaning of this result is not yet clear to us.

Let us comment now on the uniqueness of $AdS_5$ extensions of the
super-Poincar\'e algebra with central charges. For $N$-extended supersymmetry,
$OSp(1/8N)$ is the unique extension of the super-Poincar\'e algebra
in Eq.~(\ref{2}) with the following properties:
a) all right-hand sides of the
fermionic anticommutators are nonzero, and form a
simple Lie algebra; b) it contains the group $O(4,2)\times U(N)$;
c) it has the same number of fermionic charges as the superconformal
algebra $U(2,2/N)$. This follows from the classification of all
superalgebras based on simple Lie algebras given in ref.~\cite{nrs}.
Likewise, the ``intermediate'' algebra $U(2N,2N/1)$ is the
unique superconformal extension of algebra~(\ref{2}) where all chiral
Poincar\'e anticommutators are set to zero. For $N=1$ one can say more. In that
case, indeed, it was shown
in ref.~\cite{vv} that only two extensions of the super-Poincar\'e algebra
exist. One, in which all central charges are set to zero,
is the usual $U(2,2/1)$; the other is, necessarily, $OSp(1/8)$. Notice that
in this case the ``intemediate'' algebra is also $U(2,2,/1)$. In this case
${\cal SP}^+$ is just the $N=1$ supertranslational algebra considered
in~\cite{ds,w3}, while~\cite{gil}
\beq
{\cal L}^0 = R\times SO(3,3),
\eeq{last}
since $SL(4,R)\sim SO(3,3)$.
The standard Lorentz transformations and $U(1)$ R-symmetry
correspond to the subgroup $SO(3,1)\times SO(2)\subset SO(3,3)$.

Conformal extensions exist also in lower dimensions. In $AdS_4$, $OSp(N/4,R)$
can be enlarged to $OSp(1/4N,R)$. Here the embedding of the
$O(3,2)\sim Sp(4,R)$ isometry of $AdS_4$ is
\beq
Sp(4N,R)\rightarrow O(N)\times Sp(4,R),
\eeq{f1}
Similarly to the discussion in footnote 4.

The 3-grading structure is
\bea
{\cal L} &=& {\cal SP}^+ + {\cal L}^0 + {\cal SC}^-, \nonumber \\
{\cal L}^0 &=& R\times SL(2N,R) \rightarrow R\times SL(2,R)\times O(N).
\eea{f2}
Here $SL(2,R)\sim O(2,1)$ is the 3-$D$ Lorentz group, and ${\cal SP}^+$ is
the maximal central extension of the 3-$D$ supertranslation algebra in
ref.~\cite{fp}.

Now, let us discuss 1/2 BPS states for $N=8$. These BPS brane configurations
were discussed in~\cite{akly}. For this purpose, it is convenient to write
the ${\cal SP}^+$ algebra using a triality change of basis, where
$Q_{\alpha\, i}$ is a chiral spinor representation of $SO(8)$:
\bea
\{ Q_{\alpha\, i}, Q_{\beta\, j}\} &=& (\gamma^\mu C)_{\alpha\beta}P_\mu
\delta_{ij} + (\gamma^\mu C)_{\alpha\beta}\gamma_{ij}^{[ABCD]}Z^+_{\mu\,
[ABCD]} + C_{\alpha\beta}\gamma_{ij}^{[AB]}Z_{[AB]}, \nonumber \\
&& \alpha,\beta=1,2,\;\;\; i,j=1,..,8,\;\;\; A,B=1,.,8.
\eea{f3}
$Z^+_{\mu\,[ABCD]}$ is $O(8)$ self-dual.
For zero-branes, $Z_{[AB]}$ can be brought to a normal form by an $SO(8)$
rotation. 1/2 BPS states correspond to 3 vanishing eigenvalues of $Z_{[AB]}$,
that break the $O(8)$ R-symmetry as
\beq
O(8)\rightarrow O(6)\times O(2).
\eeq{f4}
For strings, the matrix $\gamma_{ij}^{[ABCD]}Z^+_{\mu\,[ABCD]}$ saturates
the 1/2 BPS bound  when $Z^+_{\mu\,[ABCD]}$ is a singlet under
$O(4)\times O(4)\subset O(8)$. These breaking patterns seem to agree with
the explicit analysis of ref.~\cite{akly}.

Finally, let us comment about other symplectic extensions.
An interesting embedding, corresponding to a $D5/D3$ brane system, is
$OSp(8N,R)\rightarrow O^*(4)\times USp(2N)$, where $O^*(4)\sim O(2,1)\times
O(3)$. This is an extension of the $OSp(4^*/2N)$ superalgebra, and
it was considered in ref.~\cite{cgmp}.

The analysis performed in this paper was made without enlarging the fermionic
sector of the superalgebra. If one doubles the spinor charges, instead, other
extensions such as $OSp(2/32,R)$, $OSp(1/32,R)_L\times OSp(1/32,R)_R$, or
$OSp(1/64,R)$ are possible. The two latter algebras, have been proposed
recently as minimal superalgebras for describing an $AdS_{11}$ version of
M-theory~\cite{h,g}, or to unify space-time symmetries in extra
dimensions~\cite{bars}.
\vskip .1in
\noindent
{\bf Acknowledgements}
\vskip .1in
\noindent
We would like to thank R. Stora and A. Zaffaroni for useful discussions and
comments.
M.P. would like to thank the Scuola Normale Superiore, Pisa, Italy,
for its kind hospitality during completion of this paper.
M.P. is supported in part by NSF grant no. PHY-9722083.
S.F. is supported in part by the EEC under TMR contract ERBFMRX-CT96-0090, ECC
Science Program SCI$^*$-CI92-0789 (INFN-Frascati), DOE grant
DE-FG03-91ER40662.


\begin{thebibliography}{66666}
\bibitem{n} W. Nahm, Nucl. Phys. B135 (1978) 149.
\bibitem{hls} R. Haag, J. T. Lopuszanski and  M. Sohnius, Nucl. Phys. B88
(1975) 257.
\bibitem{agit} J.A. de Azcarraga, J.P. Gauntlett, J.M. Izquierdo and P.K.
Townsend, Phys. Rev. Lett. 63 (1989) 2443.
\bibitem{t1} P.K. Townsend, published in {\em Proceedings of PASCOS/Hopkins
1995}, hep-th/9507048.
\bibitem{b} I. Bars, Phys. Rev. D54 (1996) 5203, hep-th/9604139.
\bibitem{t2} P.K. Townsend, Nucl. Phys. Proc. Suppl. 68 (1998) 11,
hep-th/9708034.
\bibitem{fp} S. Ferrara and M. Porrati, Phys. Lett. B423 (1998) 255,
hep-th/9711116.
\bibitem{h} P. Horava, Phys. Rev. D59 (1999) 046004, hep-th/9712130.
\bibitem{g} M. G\"unaydin, Nucl. Phys. B528 (1998) 432, hep-th/9803138.
\bibitem{gil} R. Gilmore, {\em Lie groups, Lie algebras, and some of their
applications,} Wiley-Interscience, New York, NY (1974).
\bibitem{cgmp} B. Craps, J. Gomis, D. Mateos and A. Van Proeyen,
hep-th/9901060.
\bibitem{crem} E. Cremmer, in {\em Superspace and Supergravity}, S.W. Hawking
and M. Ro\v{c}ek Eds., Cambridge U. Press, Cambridge, UK (1981).
\bibitem{g2} M. G\"unaydin, Nuovo Cim. 29A (1975) 467.
\bibitem{g3} I. Bars and M. G\"unaydin, Comm. Math. Phys. 91 (1983) 31.
\bibitem{w2} E. Witten, JHEP 9807 (1998) 006, hep-th/9805112;
S. Gukov, M. Rangamani and E. Witten, JHEP 9812 (1998) 025, hep-th/9811048.
\bibitem{malda} J. Maldacena, Adv. Theor. Math. Phys. 2 (1998) 231,
hep-th/9711200.
\bibitem{gkp} S.S. Gubser, I.R. Klebanov and
A.M. Polyakov, Phys.Lett. B428 (1998) 105, hep-th/9802109.
\bibitem{w1} E. Witten, Adv. Theor. Math. Phys. 2 (1998) 253, hep-th/9802150.
\bibitem{zu} B. Zumino, J. Math. Phys. 3 (1962) 1055.
\bibitem{kvp} P. Claus, R. Kallosh, A. Van Proeyen, Nucl. Phys. B518 (1998)
150, hep-th/9711161.
\bibitem{lm} J.T. Liu and R. Minasian, hep-th/9903269.
\bibitem{aky} C. Ahn, H. Kim and H.S. Yang, Phys. Rev. D59 (1999)
106002, hep-th/9808182.
\bibitem{cm} S. Coleman and J. Mandula, Phys. Rev. 159 (1967) 1251.
\bibitem{nrs} W. Nahm, V. Rittenberg and M. Scheunert, J. Math. Phys. 17
(1976) 1626, 1640; V.G. Kac, Comm. Math. Phys. 53 (1977) 31.
\bibitem{vv} J.W. van Holten and A. Van Proeyen, J. Phys. A15 (1982)
3763.
\bibitem{ds} G. Dvali and M. Shifman, Phys. Lett. B396 (1997) 64,
Erratum-ibid. B407 (1997) 452, hep-th/9612128.
\bibitem{w3} E. Witten, Nucl. Phys. B507 (1997) 658, hep-th/9706109.
\bibitem{akly} C. Ahn, H. Kim, B.-H. Lee and H.S. Yang, hep-th/9811010.
\bibitem{bars} I. Bars, C. Deliduman and D. Minic, hep-th/9904063.
\end{thebibliography}
\end{document}